\documentclass[sigconf, nonacm]{acmart}

\usepackage[T1]{fontenc}
\usepackage[utf8]{inputenc}
\usepackage{enumitem}
\usepackage{textcomp}

\usepackage{algorithm,algpseudocode}
\usepackage{amsmath,amsfonts,amsthm}
\usepackage{siunitx}
\usepackage{pifont,circledsteps}

\usepackage{graphicx,svg,subcaption}
\usepackage{tikz,circuitikz,pgfplots}
\usepackage{booktabs,multirow,multicol}
\usepackage{xcolor}
\usepackage{dblfloatfix}

\usepgfplotslibrary{groupplots, statistics}
\usetikzlibrary{arrows.meta}
\pgfplotsset{compat=1.18}

\newcommand{\checkmarkx}{\textcolor{ACMGreen}{\ding{51}}}
\newcommand{\crossmarkx}{\textcolor{ACMRed}{\ding{55}}}

\setcopyright{none} 
  \thanks{Distribution Statement A: Approved for public release. Distribution is unlimited.}

\title[GainSight: A Unified Framework for Data Lifetime Profiling and Heterogeneous Memory Composition]{GainSight: A Unified Framework for Data Lifetime Profiling\\ and Heterogeneous Memory Composition}

\author{Peijing Li}
\orcid{0000-0001-8517-6109}
\email{peli@stanford.edu}
\affiliation{
  \institution{Stanford University}
  \city{Stanford}
  \state{CA}
  \country{USA}
}
\author{Matthew Hung}
\email{mathu@stanford.edu}
\affiliation{
  \institution{Stanford University}
  \city{Stanford}
  \state{CA}
  \country{USA}
}
\author{Yiming Tan}
\email{yimingt@stanford.edu}
\affiliation{
  \institution{Stanford University}
  \city{Stanford}
  \state{CA}
  \country{USA}
}
\author{Konstantin Ho\ss feld}
\email{hossfeld@stanford.edu}
\orcid{0009-0005-9542-3317}
\affiliation{
  \institution{Stanford University}
  \city{Stanford}
  \state{CA}
  \country{USA}
}
\author{Jake Cheng Jiajun}
\email{jiajunc@stanford.edu}
\affiliation{
  \institution{Stanford University}
  \city{Stanford}
  \state{CA}
  \country{USA}
}
\author{Shuhan Liu}
\orcid{0000-0001-5705-565X}
\email{shliu98@stanford.edu}
\affiliation{
  \institution{Stanford University}
  \city{Stanford}
  \state{CA}
  \country{USA}
}
\author{Lixian Yan}
\orcid{0000-0003-3992-0157}
\email{lxyn5869@stanford.edu}
\affiliation{
  \institution{Stanford University}
  \city{Stanford}
  \state{CA}
  \country{USA}
}
\author{Xinxin Wang}
\orcid{0000-0003-0136-5936}
\email{xxwang1@stanford.edu}
\affiliation{
  \institution{Stanford University}
  \city{Stanford}
  \state{CA}
  \country{USA}
}
\author{Philip Levis}
\orcid{0000-0003-2934-2701}
\email{pal@cs.stanford.edu}
\affiliation{
  \institution{Stanford University}
  \city{Stanford}
  \state{CA}
  \country{USA}
}
\author{H.-S. Philip Wong}
\orcid{0000-0002-0096-1472}
\email{hspwong@stanford.edu}
\affiliation{
  \institution{Stanford University}
  \city{Stanford}
  \state{CA}
  \country{USA}
}
\author{Thierry Tambe}
\orcid{0000-0002-6411-9620}
\email{ttambe@stanford.edu}
\affiliation{
  \institution{Stanford University}
  \city{Stanford}
  \state{CA}
  \country{USA}
}

\begin{document}
  \theoremstyle{acmdefinition}
\newtheorem{takeaway}[theorem]{Takeaway}

\begin{abstract}
  As AI workloads drive increasing memory requirements, domain-specific accelerators need higher-density on-chip memory beyond what current SRAM scaling trends can provide.
  Simultaneously, the vast amounts of short-lived data in these workloads make SRAM overprovisioned in retention capability.
  To address this mismatch, we propose a wholesale shift from uniform SRAM arrays to heterogeneous on-chip memory, incorporating denser short-term RAM (StRAM) devices whose limited retention times align with transient data lifetimes.

  To facilitate this shift, we introduce GainSight, the first comprehensive, open-source framework that aligns dynamic, fine-grained workload lifetime profiles with memory device characteristics to enable generation of optimal StRAM memory compositions.
  GainSight combines retargetable profiling backends with an architecture-agnostic analytical frontend.
  The various backends capture cycle-accurate data lifetimes, while the frontend correlates workload patterns with StRAM retention properties to generate optimal memory compositions and project performance.

  GainSight elevates data lifetime to a first-class design consideration for next-generation AI accelerators, enabling systematic exploitation of data transience for improved on-chip memory density and efficiency.
  Applying GainSight to MLPerf Inference and PolyBench workloads reveals that 64.3\% of first-level GPU cache accesses and 79.01\% of systolic array scratchpad accesses exhibit sub-microsecond lifetimes suitable for high-density StRAM, with optimal heterogeneous on-chip memory compositions achieving up to 3\texttimes{} active energy and 4\texttimes{} area reductions compared to uniform SRAM hierarchies.

  To facilitate adoption and further research, GainSight is open-sourced at \url{https://gainsight.stanford.edu/}.
  \end{abstract}

  \maketitle

\section{Introduction}\label{sec:intro}

The rapid growth of data-intensive AI workloads drives demand for greater memory capacity in domain-specific accelerators to optimize locality and energy efficiency.
Large language models (LLMs) exemplify this demand, with parameter size scaling by almost 400\texttimes{} every two years while accelerator memory capacity only doubles in that same span~\cite{gholami_ai_2024}.
This widening gap exacerbates the proverbial \textit{memory wall}, where insufficient on-chip memory capacity creates performance bottlenecks and forces costly off-chip memory accesses~\cite{asad_survey_2022}.

\begin{figure}[t!]
  \centering
  \includegraphics[width=0.8\linewidth]{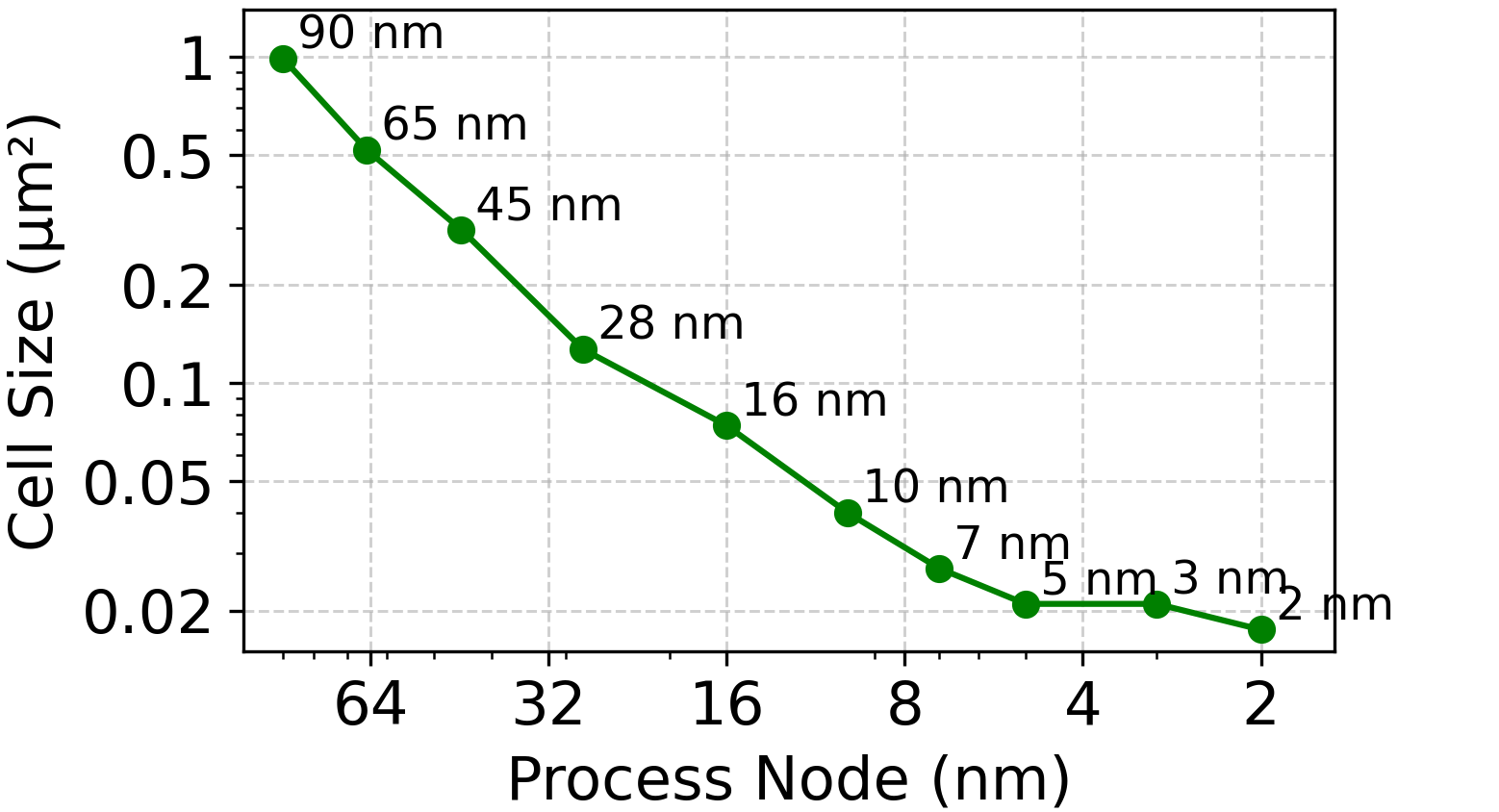}
  \caption{Scaling trend of 6T SRAM, revealing a stagnating trajectory~\cite{yeap_2nm_2024, zhang_11_2024}.}
  \label{fig:sramscaling}
\end{figure}

The conventional approach of increasing on-chip SRAM capacity faces significant scaling limitations, as shown in Figure~\ref{fig:sramscaling}.
On-chip SRAM scaling has plateaued at \SI{0.021}{\micro\meter\squared} per cell across \qtyrange{5}{3}{\nano\meter} process nodes~\cite{zhang_11_2024}.
Even the latest gate-all-around (GAA) 2nm transistor process struggles to restore earlier scaling trajectories~\cite{yeap_2nm_2024}, as maintaining the delicate variation control between transistor sizing and threshold voltages becomes increasingly difficult at smaller geometries and lower operating voltages~\cite{gul_sram_2022}.
This scaling challenge creates a critical \unit[per-mode = symbol]{perf\per\milli\meter\squared} bottleneck for accelerator design as integrating more SRAM on reticle-limited dies reduces
the available area for logic, forcing difficult trade-offs between compute and on-chip memory resources.

At the same time, data lifetime -- empirically defined as the duration that a data structure is retained in memory before it is no longer needed -- is highly heterogeneous within and across workloads and hardware architectures.
This exposes a core design tension: \textit{while existing memory systems offer consistent and uniform data retention capabilities, real-world data usage patterns require widely different lifetime durations}.

Specifically, data short-livedness is increasingly pervasive in modern AI workloads.
AI analytic pipelines commonly write, quickly ingest, and then discard or overwrite gigabytes of information.
To list only three examples in LLM inference: (1) the decoding process commonly generates multiple candidate output tokens only to retain one;
(2) massive attention score matrices exist just long enough for softmax computation before invalidation;
(3) particularly in high-throughput auto-regressive LLM inference, the \textit{key} and \textit{value} activation matrices, known as the KV cache, monopolize GPU HBM and are temporarily moved on-chip into SRAM before immediate consumption by the tensor cores.
We can back-calculate that many of these generated data structures persist for mere microseconds in modern hardware accelerators or GPUs.

These large amounts of highly transient data represent an architectural opportunity: rather than store them in SRAM, which dominates on-chip area, they can be stored in higher-density, short-retention memory technologies.

Given these trends, conventional memory can evolve from a uniform address space of random access memory to a heterogeneous collection of denser short-term and long-term memories with varying retention properties that match application-specific traffic patterns and lifetime requirements~\cite{liu_future_2024}.
This transformation mirrors the evolution from general-purpose CPUs to domain-specific accelerators in the computing space, where specialization around workload characteristics enabled greater performance and efficiency.
Just as compute has specialized, so too must memory.

Table~\ref{tab:memory_types} presents various memory technologies for on-chip integration, with a particular focus on short-term RAM (StRAM) devices
\footnote{The term ``device'' in this paper refers to memory cells of a particular memory technology (e.g., SRAM, GCRAM); this should not be confused with system-level I/O ``devices'' such as GPUs, network cards, etc.}
that directly address the data transience challenge.
StRAM devices like gain cell RAM (GCRAM) and embedded DRAM (eDRAM) provide 2-4\texttimes{} higher density and significantly lower leakage power than SRAM for transient data, despite \si{\micro\second}-scale retention times~\cite{bonetti_gain-cell_2020,giterman_1-mbit_2020,liu_gain_2023,liu_design_2024,harel_16-kb_2024}.
The key insight is that these retention constraints become advantages rather than limitations when data lifetimes are shorter than device retention times, enabling refresh-free operation with superior density and lower energy consumption.

\begin{table}[tb]
    \centering
    \footnotesize
    \caption{Comparison of on-chip memory technologies: from traditional SRAM to short-term RAM (StRAM) offering higher density and lower read/write active energies by trading off retention time. Long-term RAM (LtRAM) options are also included for completeness.}\label{tab:memory_types}
    \resizebox{\columnwidth}{!}{
  \begin{tabular}{llll}
    \toprule
    \multicolumn{1}{c}{} &
    \multicolumn{1}{c}{\textbf{SRAM}} &
    \multicolumn{1}{c}{\textbf{Long-term RAM}} &
    \multicolumn{1}{c}{\textbf{Short-term RAM}} \\ \midrule
    \textbf{Structure} & 6T SRAM & \begin{tabular}[c]{@{}l@{}}MRAM, RRAM,\\FeRAM, etc.\end{tabular} & \begin{tabular}[c]{@{}l@{}}2T/3T GCRAM,\\1T1C eDRAM\end{tabular} \\ \midrule
    \textbf{Benefits} & \begin{tabular}[c]{@{}l@{}}Long retention \&\\balanced R/W perf.\end{tabular} & \begin{tabular}[c]{@{}l@{}}Dense, 3D-stackable,\\long retention,\\low read energy\end{tabular} & \begin{tabular}[c]{@{}l@{}}Dense, 3D-stackable,\\low leakage power\end{tabular} \\ \midrule
    \textbf{Drawbacks} & \begin{tabular}[c]{@{}l@{}}Area scaling issues,\\high static power\end{tabular} & \begin{tabular}[c]{@{}l@{}}Expensive writes,\\limited endurance\end{tabular} & \begin{tabular}[c]{@{}l@{}}Short retention,\\expensive refreshes\end{tabular} \\ \midrule
    \textbf{Uses} & \begin{tabular}[c]{@{}l@{}}Fast read and\\write caching\end{tabular} & \begin{tabular}[c]{@{}l@{}}Rarely written,\\static data cache\end{tabular} & \begin{tabular}[c]{@{}l@{}}Fast write-then-read\\ops. for dynamic data\end{tabular} \\ \bottomrule
  \end{tabular}
}

\end{table}

\begin{table*}[t]
  \caption{Comparison of on-chip memory runtime profilers}
  \label{tab:profilers}
  \resizebox{\linewidth}{!}{
  \centering
  \begin{tabular}{lcccccccc}
    \toprule
    \multicolumn{1}{c}{\textbf{Profiler}} & \textbf{\begin{tabular}[c]{@{}c@{}}GPU\\ Profiling\end{tabular}} & \textbf{\begin{tabular}[c]{@{}c@{}}Accelerator\\ Profiling\end{tabular}} & \textbf{\begin{tabular}[c]{@{}c@{}}Profile \\ Kernels\end{tabular}} & \textbf{\begin{tabular}[c]{@{}c@{}}Profile \\ Cache Lines\end{tabular}} & \textbf{\begin{tabular}[c]{@{}c@{}}Profile Byte\\ Accesses\end{tabular}} & \textbf{\begin{tabular}[c]{@{}c@{}}Correlate with\\ Mem. Device\end{tabular}} & \textbf{\begin{tabular}[c]{@{}c@{}}Auto Mem. Array\\ Composition\end{tabular}} & \textbf{\begin{tabular}[c]{@{}c@{}}Open-\\ Source\end{tabular}} \\
    \midrule
    \textbf{MemSpy}~\cite{memspy} & \crossmarkx & \crossmarkx & \checkmarkx & \checkmarkx & \crossmarkx & \crossmarkx & \crossmarkx & \crossmarkx \\
    \textbf{CacheGrind}~\cite{valgrind-cachegrind} & \crossmarkx & \crossmarkx & \crossmarkx & \checkmarkx & \crossmarkx & \crossmarkx & \crossmarkx & \checkmarkx \\
    \textbf{perf mem}~\cite{linux_perf} & \crossmarkx & \crossmarkx & \checkmarkx & \checkmarkx & \checkmarkx & \crossmarkx & \crossmarkx & \checkmarkx \\
    \textbf{DProf}~\cite{pesterev_locating_2010} & \crossmarkx & \crossmarkx & \checkmarkx & \crossmarkx & \crossmarkx & \crossmarkx & \crossmarkx & \checkmarkx \\
    \textbf{Nsight Compute}~\cite{nvidia_corporation_nsight_2024} & \checkmarkx & \crossmarkx & \checkmarkx & \crossmarkx & \crossmarkx & \crossmarkx & \crossmarkx & \crossmarkx \\
    \textbf{rocProfiler}~\cite{rocprofiler} & \checkmarkx & \crossmarkx & \checkmarkx & \crossmarkx & \crossmarkx & \crossmarkx & \crossmarkx & \checkmarkx \\
    \textbf{GMProf}~\cite{GMProf} & \checkmarkx & \crossmarkx & \checkmarkx & \checkmarkx & \crossmarkx & \crossmarkx & \crossmarkx & \checkmarkx \\
    \textbf{GainSight (Ours)} & \checkmarkx & \checkmarkx & \checkmarkx & \checkmarkx & \checkmarkx & \checkmarkx & \checkmarkx & \checkmarkx \\
    \bottomrule
  \end{tabular}
  }
\end{table*}

We therefore reimagine on-chip memory through differentiated compositions of GCRAM, eDRAM, and SRAM that exploit data transience to outperform traditional SRAM-centric designs.
Realizing these gains requires treating data lifetime as a first-class design constraint and developing systematic methodologies and tools to correlate lifetime patterns with memory device retention properties.

However, existing accelerator design workflows and workload profiling tools are insufficient for providing this level of insight.
Prior methodologies for designing memory arrays with StRAM devices~\cite{chen_dadiannao_2014,tu_rana_2018} relied on static dataflow heuristics without profiling of transient lifetime patterns, rendering them less extensible to workloads exhibiting complex dynamic memory access patterns.
Existing profilers~\cite{GMProf,nvidia_corporation_cupti_2025,pesterev_locating_2010} provide coarse-grained performance counters that fail to capture the fine-grained, cycle-accurate data lifetime dynamics necessary for StRAM evaluation.
Their architecture-specific implementations also prevent systematic analysis across various accelerator backends beyond CPUs and GPUs~\cite{samajdar2020systematic}.

These deficiencies leave critical questions about accelerator workloads and prospective memory architectures unanswered:
\begin{itemize}
    \item How long do data objects persist in on-chip memory across different workloads and architectures?
    \item How do lifetime patterns correlate with StRAM retention limits to minimize refresh overhead?
    \item Given a library of StRAM devices, how can we compose them in an on-chip memory array to create an optimal mix that minimizes area and active energy consumption?
\end{itemize}

To address these research questions, we introduce \textit{GainSight}, the first comprehensive framework for lifetime-aware heterogeneous on-chip memory design that establishes data lifetime as a systematic design discipline.
Unlike existing profilers that provide narrow, platform-specific cache statistics, GainSight pioneers a unified methodology that bridges dynamic workload lifetime patterns with StRAM device characteristics to identify refresh-free operation opportunities.
This work makes three synergistic contributions to lifetime-aware on-chip memory design:

\begin{itemize}
    \item \textbf{Profile-Guided Heterogeneous Memory Composition:} We establish the first systematic approach for automatically generating optimal heterogeneous on-chip memory compositions based on dynamic workload data lifetimes and memory access patterns. This transforms memory design from intuition-driven to profile-guided optimization that exploits workload transience, providing actionable guidelines for accelerator design.
    \item \textbf{Unified Profiling and Evaluation Framework:} We implement GainSight through retargetable accelerator backends capturing fine-grained memory access traces across various accelerator architectures, coupled with an architecture-agnostic analytical frontend supporting multiple memory device evaluations. This infrastructure systematically correlates processing architectures, StRAM devices, and workloads to inform heterogeneous memory composition.
    \item \textbf{Cross-Platform Case Studies:} We demonstrate GainSight's versatility through evaluations across GPU and systolic array architectures (Section~\ref{sec:experiment}), analyzing MLPerf Inference and PolyBench workloads to provide concrete provisioning guidelines. Our studies reveal that optimal heterogeneous compositions achieve up to 3\texttimes{} active energy and 4\texttimes{} area reductions compared to uniform SRAM.
\end{itemize}

To facilitate adoption and further research, GainSight's source code is available at \url{https://code.stanford.edu/tambe-lab/gainsight}, with comprehensive documentation and interactive visualizations at \url{https://gainsight.stanford.edu/}, enabling the broader community to explore lifetime-aware on-chip memory design principles.

\section{Background and Related Works}\label{sec:related}

While prior work has explored non-SRAM on-chip memory architectures and various forms of memory profiling tools, their varied scopes and limitations mean that no single methodology systematically connects fine-grained workload data lifetime profiles with the design process of memory systems in accelerators.
This section establishes the technical foundation for GainSight's methodology, covering non-SRAM short-term memory technologies, existing on-chip memory profiling and design approaches, and the data lifetime concept that enables heterogeneous memory composition.

\subsection{Short-Term RAM Devices}\label{sec:gain_cell_ram}

\begin{figure}[t!]
    \centering
    \includegraphics[width=0.95\linewidth]{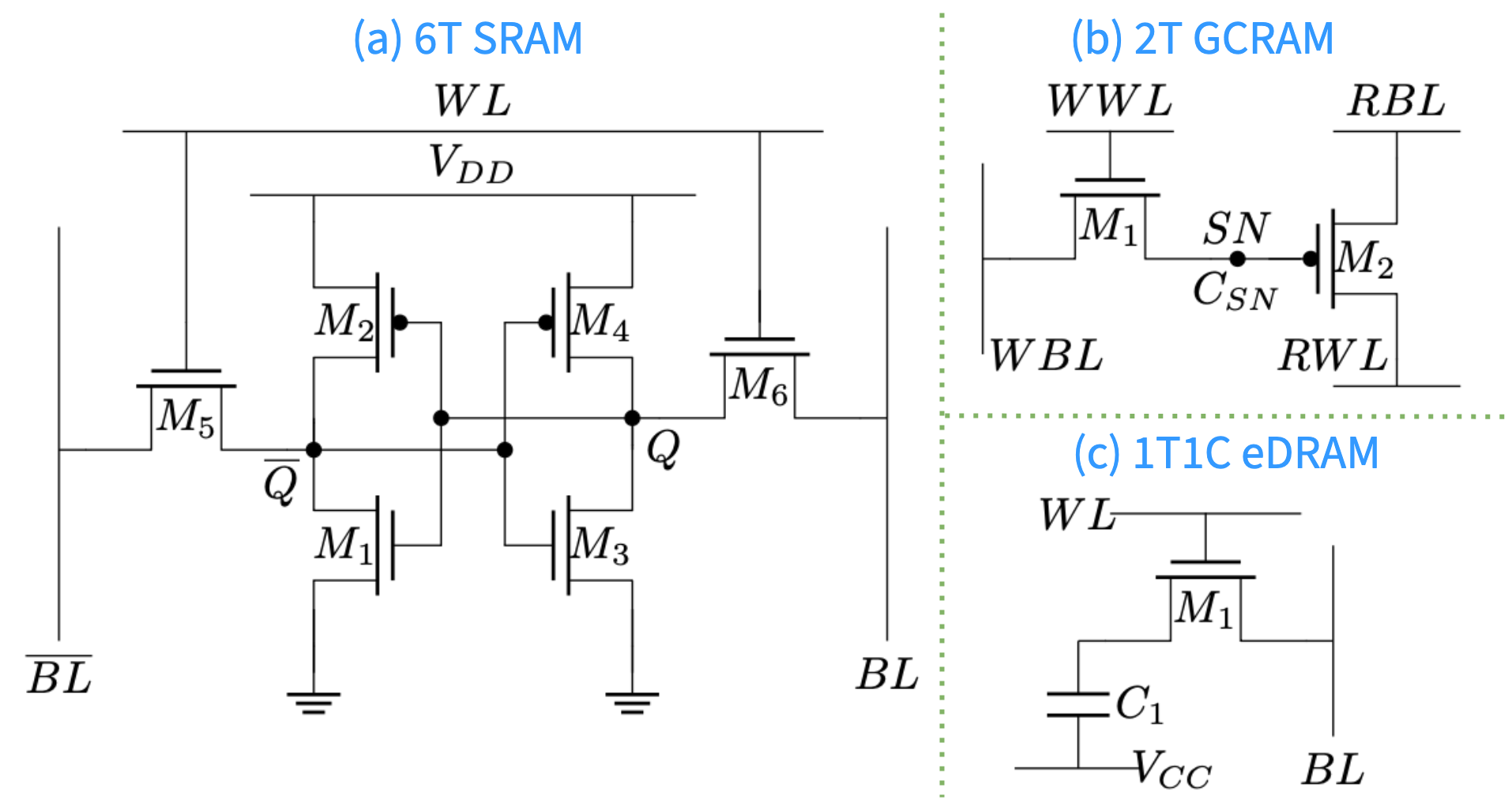}
    \caption{
        Circuit diagrams of 6T SRAM, 2T GCRAM, and 1T1C eDRAM.
        Note the smaller transistor count of GCRAM and eDRAM vs. SRAM.
        To further save area, the write transistor $M_1$ of GCRAM and the capacitor $C_1$ of eDRAM can be 3D stacked on top of silicon at backend-of-line~\cite{Liu2024FirstED,chiang_integration_2025} -- optimization opportunities unavailable to SRAM.
    }
    \label{fig:circuit}
\end{figure}

Given SRAM's scaling limitations and the transient nature of AI workload data, alternative short-term memory (StRAM) technologies have been designed to offer superior density and energy efficiency by trading off retention time for area and energy benefits.
Figure~\ref{fig:circuit} shows the circuit topologies of the memory devices evaluated in GainSight: six-transistor (6T) static random-access memory (SRAM), two-transistor (2T) gain cell RAM (GCRAM), and one-transistor-one-capacitor (1T1C) embedded DRAM (eDRAM).

\textbf{SRAM}, as the conventional on-chip memory device technology, provides fast access and indefinite retention through cross-coupled inverters but faces scaling limitations and overprovisioning for \si{\micro\second}-scale AI data lifetimes.

\textbf{Gain Cell RAM (GCRAM)} uses 2-3 transistors per cell and utilizes the read transistor as a transconductance amplifier to convert the stored voltage into an amplified bitline current for read sensing, achieving 2-4\texttimes{} higher density and lower leakage than SRAM~\cite{bonetti_gain-cell_2020}.
Comprised of silicon transistors fabricated with conventional fin field-effect (FinFET) processes, \textbf{Si-GCRAM} offers \si{\micro\second}-scale retention with SRAM-comparable latencies~\cite{giterman_1-mbit_2020,pentecost_nvmexplorer_2022}.

\textbf{Hybrid-GCRAM} replaces the write nMOS transistor in Si-GCRAM ($M_1$ in Figure~\ref{fig:circuit}(b)) with an oxide semiconductor transistor made of materials such as indium tin oxide (ITO) through back-end-of-line (BEOL) manufacturing processes.
This design achieves longer retention compared to Si-GCRAM and higher density through 3D stacking of the ITO transistor on top of silicon peripheral circuits~\cite{Chen_Oxide_2024,Liu2024FirstED}.
However, there is an inverse tradeoff between its retention and write frequency~\cite{liu_design_2024}, necessitating even closer examination of workload memory access patterns to avoid refresh overhead.

Traditionally, it has been challenging to integrate dynamic RAM (DRAM) into on-chip memory architectures due to capacitor fabrication complexities.
However, recently proposed \textbf{embedded DRAM (eDRAM)} designs enable on-chip integration using oxide transistors and BEOL processes, providing \si{\milli\second}-scale retention with high density~\cite{chiang_integration_2025} but higher active energy than GCRAM due to charge-sharing read mechanisms.

These technologies offer distinct retention-energy-area tradeoffs that GainSight can correlate with workload lifetime patterns to identify optimal heterogeneous compositions.

\subsection{Memory Profilers}\label{sec:related_profilers}

Effectively utilizing StRAM requires understanding fine-grained data lifetimes and memory access patterns through profiling of target workloads to inform on-chip memory system design decisions.
While promising in theory, existing approaches exhibit critical limitations that prevent principled exploitation of data transience.
Table~\ref{tab:profilers} compares existing profilers with GainSight's capabilities, highlighting their limitations in coarse granularities and limited architecture scopes.
We also observe that existing profilers are primarily retrospective analysis tools without systematic correlation of profiling results to prospective memory device characteristics, as they are not intended as design tools.

CPU profiling is well-established through tools like Linux \texttt{perf mem}~\cite{linux_perf} and full-system simulators such as MemSpy~\cite{memspy} that track fine-grained memory accesses and cache behavior.
DProf~\cite{pesterev_locating_2010} correlates sampled cache misses with program structures, identifying inefficiencies in data layout and cache utilization.
GPU profilers such as NVIDIA's Nsight Compute~\cite{nvidia_corporation_nsight_2024} and CUPTI~\cite{nvidia_corporation_cupti_2025} provide only kernel-level granularity and aggregated statistics regarding on-chip cache usage, without characterizing individual memory access behaviors or data lifetime patterns crucial for heterogeneous memory composition.
GMProf~\cite{GMProf} achieved finer granularity but uses abstract memory constructs that do not map to physical on-chip memory arrays.

\subsection{Accelerators with Emerging Memories}\label{sec:rel_arch}

\begin{table}[t!]
    \centering
    \caption{Architectures leveraging non-SRAM short- or long-term on-chip memory devices}
    \label{tab:devices}
    \resizebox{\columnwidth}{!}{
    \begin{tabular}{llll}
        \toprule
        \multicolumn{1}{c}{\textbf{Name}} & \multicolumn{1}{c}{\textbf{\begin{tabular}[c]{@{}c@{}}Target \\ Workload\end{tabular}}} & \multicolumn{1}{c}{\textbf{\begin{tabular}[c]{@{}c@{}}Memory \\ Type\end{tabular}}} & \multicolumn{1}{c}{\textbf{\begin{tabular}[c]{@{}c@{}}Memory\\ Device\end{tabular}}} \\ \midrule
        \textbf{DaDianNao}~\cite{chen_dadiannao_2014} & CNN inference & Short-term & eDRAM \\
        \textbf{RANA}~\cite{tu_rana_2018} & CNN inference & Short-term & eDRAM \\
        \textbf{CAMEL}~\cite{zhang_camel_2024} & CNN/Transformer training & Short-term & GCRAM \\
        \textbf{STT-AI Ultra}~\cite{mishty_designing_2021} & CNN training/inference & Long-term & MRAM \\
        \textbf{CHIMERA}~\cite{giordano_chimera_2021} & CNN training & Long-term & RRAM \\ \bottomrule
    \end{tabular}
    }
\end{table}

Various attempts have been made, as summarized in Table~\ref{tab:devices}, to address \textit{Memory Wall} bottlenecks in accelerators by integrating non-SRAM on-chip memory devices as shown in Table~\ref{tab:memory_types}, including gain cell RAM (GCRAM), embedded DRAM (eDRAM), resistive RAM (RRAM) and magnetoresistive RAM (MRAM).

The common thread in these designs is that they combine conventional processing elements (e.g., SIMD, systolic arrays) with novel memory devices, typically using static heuristics of predictable dataflows in workloads to estimate data lifetimes relative to device retention characteristics.
However, these heuristic-based approaches are typically specialized and lack generalization, and must be re-evaluated when newer and emerging models such as state-space models~\cite{gu_mamba_2024} come online.

GainSight bridges the aforementioned research gaps by providing comprehensive fine-grained profiling across arbitrary workloads and hardware backends, as well as the ability to systematically correlate these profiles with StRAM device characteristics to generate optimal heterogeneous memory compositions.
This unified approach resolves both profiling granularity limitations and the lack of quantitative design insights in current memory systems.
Such systematic design exploration enables next-generation heterogeneous memory architectures that effectively exploit workload data transience.

\subsection{Data Lifetime Concept}\label{sec:lifetime}

Data lifetime quantifies the time interval during which a data structure incorruptibly lives in memory to satisfy application functional correctness, forming the key metric for correlating workload behavior with retention-constrained memory devices.

\begin{definition}
    \label{def:lifetime}
    The \textbf{lifetime} of a data structure at a given memory address is the time between its \textbf{first write} and \textbf{last read} before being overwritten or invalidated.
\end{definition}

This definition adapts across memory implementations.
Scratchpad lifetimes span from stores to final reads before overwriting, while cache lifetimes extend from stores or misses to the last hit before subsequent misses or stores.

Memory device suitability depends on lifetime-retention correlation. \textbf{Short-lived} data with lifetimes below device retention enables refresh-free operation, while \textbf{long-lived} data requires refresh mechanisms or longer-retention alternatives.

Figure~\ref{fig:retention} demonstrates this principle using Llama-3.2-11B H100 kernel lifetimes~\cite{grattafiori_llama_2024} against Si-GCRAM and Hybrid-GCRAM characteristics, with write frequency effects on GCRAM retention capabilities.

This analysis reveals that GEMM operations exhibit lifetimes compatible with the retention capabilities of Si-GCRAM and tensor operations align with Hybrid-GCRAM, while normalization operations' lifetimes exceed both memory devices' retention limits and require eDRAM or SRAM to avoid refresh overhead.
These patterns exemplify the foundation for GainSight's systematic approach to matching data transience with appropriate device technologies in heterogeneous memory compositions to enable refresh-free operation.

\begin{figure}[t]
    \centering
    \begin{tikzpicture}[]
  \begin{axis}[
    width=\columnwidth,
    height=0.5\columnwidth,
    xlabel={Write Frequency (\si{\mega\hertz})},
    ylabel={Lifetime (\si{\micro\second})},
    xlabel near ticks,
    ylabel near ticks,
    xmin=0.07, xmax=2.38,
    ymin=5e-7, ymax=2.62e-5,
    ymode=log,
    xtick={0.5, 1, 1.5, 2, 2.5},
    xticklabels={$500$, $1000$, $1500$, $2000$, $2500$},
    ytick={5e-7, 1e-6, 2e-6, 4e-6, 8e-6, 1.6e-5},
    yticklabels={0.5, 1, 2, 4, 8, 16},
    scaled y ticks=false,
    grid=both,
    grid style={line width=.1pt, draw=gray!10},
    major grid style={line width=.2pt,draw=gray!50},
  ]
  
    \addplot[
      color=blue,
      mark=*,
      mark size=0pt,
      line width=1.5pt,
    ] coordinates {
      (0.068, 1.03e-5)
      (0.163, 1.03e-5)
      (0.277, 1.03e-5)
      (0.489, 1.03e-5)
      (0.709, 1.03e-5)
      (0.96,  1.03e-5)
      (1.23,  1.03e-5)
      (1.56,  1.03e-5)
      (1.87,  9.80e-6)
      (1.96,  9.22e-6)
      (2.12,  6.80e-6)
      (2.39,  1.68e-6)
    };
    
    \addplot[
      color=green!60!black,
      line width=1.5pt,
    ] coordinates {
      (0,    1e-6)
      (2.5,  1e-6)
    };
    
    \addplot[only marks, mark=*, color=orange, mark size=2pt] coordinates {(1.421,17.40e-6)};
    \node[anchor=west, color=orange] at (axis cs:1.43,17.40e-6+1e-6) {\normalsize \textbf{Normalization}};
            
    \addplot[only marks, mark=*, color=purple, mark size=2pt] coordinates {(1.189,0.736e-6)};
    \node[anchor=east, color=purple, xshift=-3pt] at (axis cs:1.21,0.67e-6) {\normalsize \textbf{GEMM}};
            
    \addplot[only marks, mark=*, color=brown, mark size=2pt] coordinates {(0.122,5.90e-6)};
    \node[anchor=west, color=brown, xshift=3pt] at (axis cs:0.105,5.40e-6) {\normalsize \textbf{Residual Connection}};

    \addplot[only marks, mark=*, color=red, mark size=2pt] coordinates {(1.143,2.06e-6)};
    \node[anchor=west, color=red] at (axis cs:1.143,2.06e-6+4e-7) {\normalsize \textbf{Tensor Transpose}};
    
    \node[anchor=south west, color=blue] at (axis cs:0.05,1e-5) {\small Hybrid-GCRAM Retention Time};
    \node[anchor=south west, color=green!60!black] at (axis cs:0.05,1e-6+2e-8) {\small Si-GCRAM Retention Time};

    \node[rotate=90, anchor=south] at (axis cs:0,6e-6) {\normalsize \si{\micro\second}};
  \end{axis}
\end{tikzpicture}
    \vspace{-0.5em}
    \caption{Data lifetimes and write frequencies for Llama-3.2-11b operations~\cite{grattafiori_llama_2024} executed on a GPU backend compared against 5nm Si-GCRAM and Hybrid-GCRAM retention limits~\cite{liu_design_2024}.}
    \label{fig:retention}
\end{figure}

\begin{figure*}[t]
  \centering
  \includegraphics[width=\textwidth]{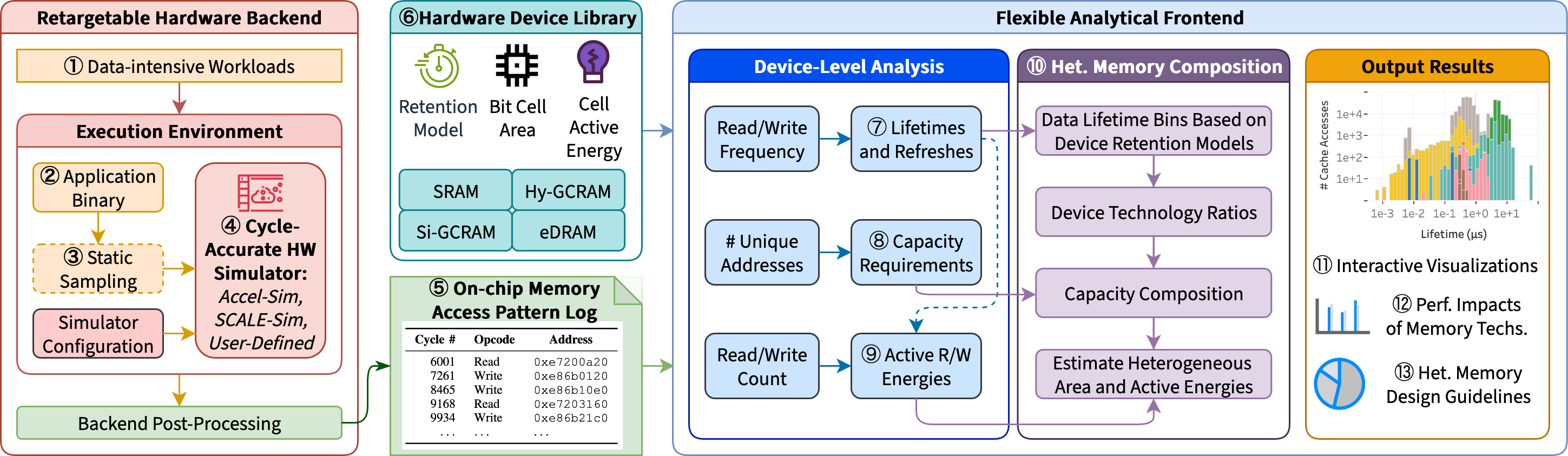}
  \caption{GainSight framework organization, comprising retargetable hardware backends and a flexible analytical frontend. The primary output of the backend is a memory access trace; the frontend inputs this trace, along with specifications of memory device models, to produce performance projections and heterogeneous on-chip memory design guidelines.}
  \label{fig:org}
\end{figure*}

\section{GainSight Organization}\label{sec:org}

GainSight is designed as a comprehensive framework for analyzing on-chip memory behavior, addressing the question: ``What are the tangible benefits of replacing on-chip SRAM with heterogeneous compositions of short-term memory devices for different workloads and hardware backends?''

To answer this, GainSight performs the following key tasks:

\begin{enumerate}
    \item Capture fine-grained on-chip memory access traces for arbitrary workloads on retargetable hardware backends.
    \item Extract lifetime statistics, read/write frequencies, and capacity utilization patterns.
    \item Correlate lifetime characteristics with memory device retention properties and other characteristics to project memory array area and active energy metrics.
    \item Generate optimal heterogeneous compositions of high-density StRAM that minimize energy consumption through refresh-free operation.
    \item Provide interactive visualizations that inform lifetime-aware design decisions.
\end{enumerate}

To meet these requirements, GainSight comprises two main components: retargetable hardware backends that connect to an architecture-agnostic analytical frontend, as shown in Figure~\ref{fig:org}.
The backends execute workloads on hardware architectural simulators and generate fine-grained, cycle-accurate memory access traces with lifetime annotations.
The frontend processes these traces to extract data lifetime statistics, correlates them with retention-limited StRAM characteristics, projects optimal heterogeneous memory compositions, and provides interactive visualizations for design insights.
The memory device library includes post-fabrication models of SRAM, Si-GCRAM~\cite{giterman_1-mbit_2020}, Hybrid-GCRAM~\cite{liu_gain_2023}, and 1T1C eDRAM~\cite{chiang_integration_2025}.

\subsection{Usage Scenario}

Hardware accelerator designers seeking to exploit data transience can leverage GainSight as a comprehensive design exploration tool that bridges workload analysis with heterogeneous memory provisioning decisions.

The primary input designers provide is a set of representative AI workloads relevant to their target application domain, which can be executed on the provided backends of a SIMD GPU or systolic array accelerator.
For specialized prospective architectures not covered by the existing backends, users may optionally implement custom backends as described in Section~\ref{sec:byob}.

GainSight generates three complementary output formats that directly inform memory hierarchy design decisions.
Quantitative JSON reports provide concrete metrics including refresh requirements, active energy consumption, and area projections for both individual StRAM devices and optimal heterogeneous compositions.
Interactive visualizations enable exploration of lifetime distributions and device correlation patterns across workloads and kernel subroutines.

Most critically, the framework produces detailed specifications for optimal heterogeneous memory compositions, including capacity quantities, area allocations, and projected energy benefits compared to uniform SRAM designs.

These outputs provide insights to enable architects to make data-driven provisioning decisions for heterogeneous memory arrays before costly physical prototyping.

The following sections describe GainSight's retargetable backends and analytical frontend architecture.

\section{Retargetable Hardware Backend}\label{sec:backend}

GainSight's retargetable hardware backends execute workloads on a variety of accelerator architectures to capture fine-grained memory access traces and data lifetime statistics.
This multi-backend approach is essential because memory access patterns and data lifetimes vary significantly across different hardware architectures, even for identical workloads, due to varying dataflow patterns and on-chip memory organizations.

Since existing runtime profilers lack the granularity required for lifetime-aware analysis as discussed in Section~\ref{sec:related_profilers}, we implement simulation-based backends by modifying architectural simulators to generate the fine-grained traces necessary for StRAM device evaluation.
Simulation-based approaches enable arbitrary memory hierarchy configurations and support evaluation of hypothetical compositions that are not yet available in physical hardware, making them essential for design space exploration of emerging memory technologies.

Our implementation includes GPU and systolic array accelerator backends, but GainSight can integrate any backend capable of generating fine-grained traces in the required format \Circled{5} in Figure~\ref{fig:org}.

\subsection{Backend 1: Simulating GPU Architectures}\label{sec:accelsim}

Single instruction, multiple data (SIMD) architectures have become the standard for accelerating deep learning workloads, with NVIDIA GPUs being a leading example.
To capture memory access patterns and data lifetime metrics for SIMD architectures, we modified Accel-Sim v1.3.0~\cite{khairy_accel-sim_2020}, a cycle-accurate GPU simulator for use as GainSight's first backend.

The GPU backend generates traces through a multi-stage process following Figure~\ref{fig:org}.
CUDA-based workloads \Circled{1} are first processed by an NVBit~\cite{villa_nvbit_2019}-based tracing utility to capture SASS assembly code for each kernel \Circled{2}.
The assembly code then undergoes offline kernel clustering and sampling \Circled{3}~\cite{avalos_baddouh_principal_2021} to select representative kernels for simulation, which is necessary due to the high computational cost of cycle-accurate GPU simulation.
These representative traces are fed into GPGPU-Sim v4.2.1~\cite{bakhoda_analyzing_2009,khairy_accel-sim_2020} to simulate program execution and collect memory access traces at each individual cycle and instruction \Circled{4}, generating detailed on-chip cache access traces to facilitate lifetime analysis \Circled{5}.

Our modifications to Accel-Sim include logging call insertions at the interfaces between the streaming multiprocessors (SMs) and the L1 and L2 caches to record address, access type, timing, and hit/miss status of each cache operation.
We focus on L1 and L2 caches, which are most relevant for AI workloads and offer the greatest optimization opportunities through heterogeneous compositions.

\subsubsection{Simulator Performance and Program Sampling}

The computational overhead of cycle-accurate GPU simulation presents a significant challenge for memory system evaluation, especially for contemporary AI workloads, with Accel-Sim requiring 6-7 orders of magnitude longer execution time than native GPU hardware~\cite{villa_need_2021}.

To address this simulation speed challenge, we leverage the inherent repetitiveness of deep learning computations, where similar layer structures recur throughout models~\cite{he_deep_2015, vaswani_attention_2017,restuccia_time-predictable_2021} and identical operations execute across different data batches~\cite{hochreiter_long_1997, brown_language_2020}, enabling us to sample representative kernels rather than simulating every single kernel in a workload.
We adopt the Principal Kernel Analysis (PKA) methodology~\cite{avalos_baddouh_principal_2021}, a silicon-validated sampling technique that identifies representative kernel subsets through hierarchical clustering based on offline analysis of kernel characteristics.

Table~\ref{tab:slowdown} demonstrates the effectiveness of PKA-based sampling across representative AI workloads, showing substantial reductions in simulation requirements while preserving sufficient accuracy for memory system analysis.
The sampling process reduces executed kernels to as few as 0.15\% of the original workload (LLaMA-3-8B), achieving speedups ranging from 10.89\texttimes{} to 412.76\texttimes{} across different model architectures.
The mean absolute error (MAE) for critical memory metrics remains bounded.

\begin{table}[t]
    \centering
    \caption{Key figures of merit of kernel sampling applied to selected AI workloads running on the GPU backend, with average baseline lifetimes, write frequencies, Si-GCRAM area requirements, and Si-GCRAM active energies estimated at around \SI{83.5}{\micro\second}, \SI{339}{\mega\hertz}, \SI{11}{\milli\meter\squared} and \SI{23.5}{\kilo\joule}, respectively.}
    \label{tab:slowdown}
    \resizebox{\columnwidth}{!}{
        \begin{tabular}{lrrrrrr}
            \toprule
            \textbf{Workload}                                                                        &
            \textbf{\begin{tabular}[c]{@{}c@{}}Sampled\\ (\%)\end{tabular}}                          &
            \textbf{\begin{tabular}[c]{@{}c@{}}Speedup\\ (\texttimes{})\end{tabular}}                &
            \textbf{\begin{tabular}[c]{@{}c@{}}Lifetime\\ MAE (\si{\micro\second})\end{tabular}}     &
            \textbf{\begin{tabular}[c]{@{}c@{}}Write Freq \\ MAE (\si{\mega\hertz})\end{tabular}}    &
            \textbf{\begin{tabular}[c]{@{}c@{}}Area MAE \\ (\si{\milli\meter\squared})\end{tabular}} &
            \textbf{\begin{tabular}[c]{@{}c@{}}Energy \\ MAE (\si{\kilo\joule})\end{tabular}}                                                        \\
            \midrule
            \texttt{bert-base-uncased}                                                               & 8.37\%  & 10.89  & 0.36 & 11.59 & 0.00 & 1.09 \\
            \texttt{llama-3-8b}                                                                      & 0.15\%  & 412.76 & 3.76 & 21.39 & 0.91 & 3.65 \\
            \texttt{resnet-50}                                                                       & 10.31\% & 60.97  & 3.73 & 10.09 & 0.00 & 1.99 \\
            \bottomrule
        \end{tabular}
    }
\end{table}

This workload sampling approach is extensible to other prospective backends, enabling acceleration of all simulation-based evaluations of GainSight.

\subsection{Backend 2: Simulating Systolic Arrays}
\label{sec:backend_sysarray}

For systolic array-based accelerators, we extend the open-source SCALE-Sim~\cite{samajdar2020systematic}, a Python-based simulator for convolution or GEMM-based DNN networks.
SCALE-Sim models three peripheral SRAM buffers -- input (\textit{ifmap}), weight (\textit{filter}), and output (\textit{ofmap}) -- and provides the essential features for our backend implementation: cycle-accurate simulations, configurable buffer sizes, adjustable computation dataflow, processing element array dimensions, and built-in memory access trace generation.

Following Figure~\ref{fig:org}, the backend starts from target workload specifications \Circled{1} and extracts binary forms of model hyperparameters and GEMM operands \Circled{2} executable on the cycle-accurate simulator \Circled{4}.
The simulator's key tunable parameters include processing element array dimensions and dataflow configuration, with SCALE-Sim supporting input-stationary, weight-stationary, or output-stationary modes that determine how data streams through processing elements and buffers.

\subsection{\textbf{B}ring \textbf{Y}our \textbf{O}wn (Hardware) \textbf{B}ackend}
\label{sec:byob}

Beyond the GPU and systolic array implementations, GainSight's architecture enables integration of arbitrary hardware backends capable of generating fine-grained memory access traces.
The analytical frontend processes any backend-generated traces that adhere to the standardized format shown in \Circled{5} of Figure~\ref{fig:org}.
This format includes detailed timestamps, address ranges, and read/write types for each memory access.
This extensibility allows hardware simulators or physical accelerators to serve as alternative backends, provided they can generate the required trace granularity and support compilation of target workloads \Circled{1} into executable binary code \Circled{2}.

Architectural differentiation across AI accelerators necessitates a flexible abstraction layer to handle varying memory organizations.
While GPUs employ hierarchical cache structures with L1 and L2 levels, systolic arrays utilize dedicated SRAM scratchpads for weights and feature maps, and other accelerators may implement entirely different memory architectures.
These organizational differences extend beyond processing elements themselves to encompass contrasting dataflow patterns, memory access mechanisms, and capacity distributions.

GainSight introduces memory ``subpartitions'' as a unifying abstraction.
Each subpartition represents a distinct memory component, such as cache levels in GPUs, scratchpad buffers in systolic arrays, or specialized storage in custom accelerators.
The frontend treats subpartitions as independent entities for lifetime extraction, frequency analysis, and performance projection, enabling consistent evaluation of heterogeneous memory compositions across different architectural paradigms while preserving their unique organizational characteristics.

\section{Analytical Frontend}\label{sec:frontend}

GainSight's analytical frontend transforms raw memory access traces from hardware backends into actionable design insights for heterogeneous memory architectures.
As shown in Figure~\ref{fig:org}, this architecture-agnostic module processes the cycle-accurate traces \Circled{5} generated by backends to extract key memory statistics.
It then correlates these workload characteristics with memory device retention properties to generate optimal memory compositions and performance projections.
The frontend's modular Python implementation represents a significant portion of GainSight's codebase and enables systematic evaluation across diverse accelerator architectures and workloads, establishing the foundation for data-driven heterogeneous on-chip memory design.

\subsection{Memory Device Library}

GainSight focuses on StRAM devices to bridge the mismatch between SRAM's excessive retention and the transient nature of AI workload data.
The memory device library \Circled{6} in Figure~\ref{fig:org} includes 6T SRAM, 3T silicon gain cell RAM (Si-GCRAM), 2T ITO-silicon hybrid gain cell RAM (Hybrid-GCRAM), and 1T1C oxide embedded DRAM (eDRAM), all characterized for the TSMC N5 (5nm) process node.

Memory cell models capture retention times (including write frequency dependence for GCRAM), bit cell area for density projections, and read/write energy per bit for active energy projections.
SRAM characteristics derive from TSMC specifications~\cite{yeap_5nm_2019,yeap_2nm_2024}, while Si-GCRAM parameters build upon fabricated arrays from Giterman et al.~\cite{giterman_1-mbit_2020} and Pentecost et al.~\cite{pentecost_nvmexplorer_2022}, scaled to the 5nm node (N5).
Hybrid-GCRAM statistics originate from Liu et al.'s post-layout simulations~\cite{liu_gain_2023,liu_design_2024} of $256 \times 256$ arrays.
Oxide-semiconductor eDRAM parameters derive from Chiang et al.~\cite{chiang_integration_2025}, with DRAM-specific operations like precharge amortized into per-bit energy estimates.

This memory device library organization allows straightforward extension to additional memory technologies by only requiring the addition of corresponding performance mockups while maintaining the frontend's modularity.

\subsection{Device-Level Memory Access Analysis}

The frontend's analysis pipeline begins by extracting memory statistics \Circled{6} from backend-generated traces \Circled{5}.
These statistics include read/write frequencies, unique addresses accessed, and total operation counts.

The module computes data lifetimes \Circled{7} as defined in Section~\ref{sec:lifetime}, tracking the temporal interval from each write operation until the data is overwritten or becomes invalidated.
This lifetime distribution is then correlated with the write frequency-dependent retention characteristics of each memory device in the library.
Refresh requirements are computed by identifying data objects whose lifetimes exceed the retention limits of the target device, quantifying the number of bit-level refresh operations required to maintain data integrity.

Capacity utilization \Circled{8} analyzes unique addresses remaining live simultaneously, scaled by memory block size.
This quantifies actual workload memory usage, which may be substantially less than available on-chip memory capacity.
Active energy consumption \Circled{9} combines weighted read, write, and refresh operations multiplied by device-specific costs, where refreshes are modeled as additional read-then-write operations.

These single-device projections establish baseline comparisons that inform subsequent composition optimization \Circled{10} and provide preliminary considerations for heterogeneous array provisioning.

\subsection{Heterogeneous Memory Composition}

The core innovation of GainSight's frontend lies in computing optimal heterogeneous memory compositions \Circled{10} that minimize refresh overhead while maximizing area and energy efficiency.
This optimization process begins by analyzing the distribution of data lifetimes across the workload and systematically assigning memory accesses to devices based on a lifetime-retention matching algorithm.

Each memory access is assigned to the device with the shortest retention time that still exceeds the data's lifetime requirement.
This assignment strategy leverages the general trend where Si-GCRAM, Hybrid-GCRAM, and eDRAM offer increasing retention times at the cost of higher active energy consumption.
By prioritizing memory devices with minimal necessary retention, the algorithm maximizes the utilization of more energy-efficient short-retention devices while ensuring refresh-free operation.

The assignment process yields the proportion of memory accesses allocated to each device type, which the frontend then correlates with total capacity requirements from \Circled{8} to determine the physical area and capacity specifications for each device in the heterogeneous composition.
The final optimization generates a comprehensive specification including total array capacity, area footprint, and aggregate energy consumption.
This establishes quantitative design guidelines for heterogeneous memory array implementation.

\begin{figure*}[htbp]
  \centering
  \includegraphics[width=\textwidth]{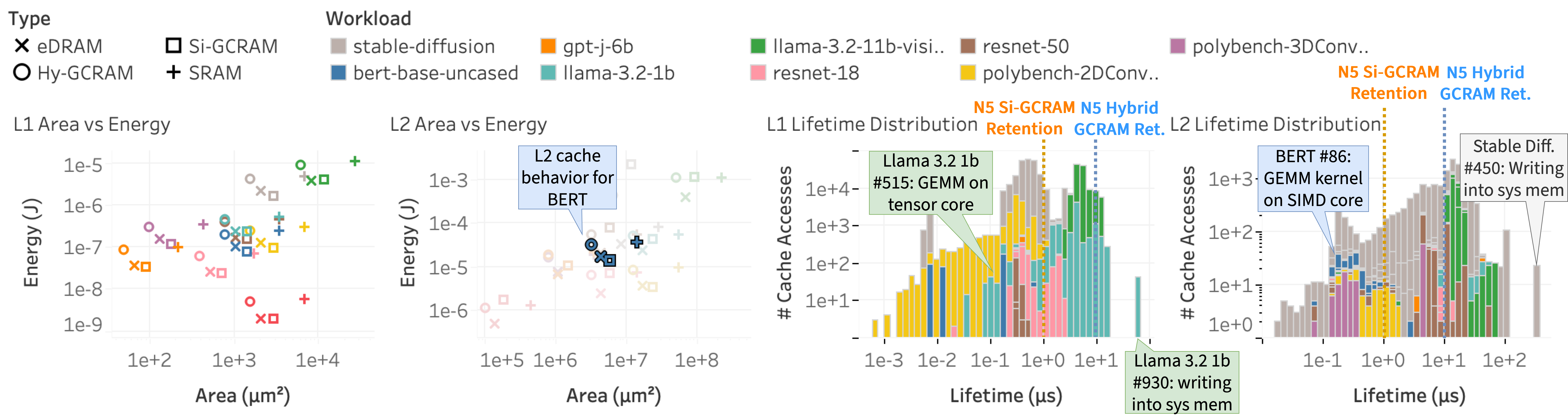}
  \caption{Energy and area projections of L1 and L2 GPU caches implemented with each memory technology, along with the data lifetime distributions of all L1 and L2 accesses across all workloads in Table~\ref{tab:gpu-workloads}.}
  \label{fig:tableau}
\end{figure*}

\subsection{Output Generation and Design Insights}\label{sec:visualization}

GainSight's frontend generates three complementary output formats to support comprehensive design analysis, as illustrated in Figure~\ref{fig:org}: interactive lifetime distribution visualizations \Circled{11}, quantitative JSON performance reports \Circled{12}, and optimal memory composition summaries \Circled{13}.

The Tableau-based visualization framework~\cite{salesforce_inc_tableau_2025} enables interactive exploration of lifetime distributions across configurations and workloads.
Figure~\ref{fig:tableau} demonstrates the interface capabilities, presenting histogram analyses of lifetime characteristics for various memory technologies and workloads.
Users can dynamically filter results and access detailed workload information through integrated tooltips, facilitating visual comparison and further design insights.

The JSON report provides quantitative metrics essential for systematic design evaluation: refresh requirements quantifying the number of bit-level refresh operations based on write frequency and retention time interactions; active energy consumption encompassing reads, writes, and refreshes; and area requirements derived from unique address access patterns.
These metrics are generated both for individual memory devices and for the computed optimal heterogeneous composition, enabling direct quantification of heterogeneous memory benefits over traditional SRAM-only designs.

The composition summary delivers concrete design specifications for the optimal heterogeneous memory configuration, including precise memory array capacity, area allocations, and energy projections.
This output directly addresses the design question regarding the quantitative area and performance impacts of heterogeneous on-chip memory design, providing actionable insights for system architects.

In the absence of physically fabricated prototypes for novel on-chip memories, these outputs provide first-order design guidance that facilitates quantitative evaluation of area, energy, and refresh trade-offs across workloads.
This capability supports critical design decisions, such as runtime memory provisioning policy development, before physical prototyping becomes available, while abstracting away detailed circuit-level effects that would require additional specialized analysis tools such as SPICE simulations.

\begin{table}[t]
  \centering
  \caption{List of workloads in GainSight's case studies.}
  \label{tab:gpu-workloads}
  \resizebox{\columnwidth}{!}{
    \begin{tabular}{lll}
      \toprule
      \textbf{Name} & \textbf{Test Suite} & \textbf{Description} \\ \midrule
      \texttt{polybench-2DConv} & PolyBench & 2D Convolution \\ \midrule
      \texttt{polybench-3DConv} & PolyBench & 3D Convolution \\ \midrule
      \texttt{llama-3.2-1b} & ML Inference & Meta's text-based LLM with 1B parameters~\cite{touvron_llama_2023} \\ \midrule
      \texttt{llama-3-8b} & ML Inference & Meta's text-based LLM with 8B parameters~\cite{touvron_llama_2023} \\
      \texttt{llama-3.2-11b-vision} & ML Inference & \begin{tabular}[c]{@{}l@{}}Meta's LLM with integrated vision adapter for image \\recognition, total of 11B parameters~\cite{touvron_llama_2023}\end{tabular} \\ \midrule
      \texttt{resnet-18} & ML Inference & CNN for image recognition with 18 layers~\cite{he_deep_2015} \\ \midrule
      \texttt{resnet-50} & ML Inference & CNN for image recognition with 50 layers~\cite{he_deep_2015} \\ \midrule
      \texttt{bert-base-uncased} & ML Inference & \begin{tabular}[c]{@{}l@{}}``Bidirectional Encoder Representation for Transformers,''\\ uncased text-based LLM with 110M parameters~\cite{devlin_bert_2019}\end{tabular} \\ \midrule
      \texttt{gpt-j-6b} & ML Inference & \begin{tabular}[c]{@{}l@{}}GPT-like text-based LLM with 6B parameters~\cite{gpt-j,mesh-transformer-jax}\\ implemented on the JAX interface\end{tabular} \\ \midrule
      \texttt{stable-diffusion} & ML Inference & Text-to-image transformer model with 3.5B parameters~\cite{esser_scaling_2024} \\ \bottomrule
    \end{tabular}
  }
\end{table}

\section{Case Study and Experiments}\label{sec:experiment}

To illustrate GainSight's practical application, we conduct two case studies targeting SIMD GPU and systolic array accelerator architectures.
These platforms were chosen for their prevalence in AI hardware and their contrasting memory organizations, providing representative evaluations of GainSight's profiling and analysis capabilities.

We executed a set of workloads from the MLPerf Inference benchmark (v5.0) suite~\cite{reddi_mlperf_2020} and PolyBench, a polyhedral benchmark of basic ML-related kernels~\cite{pouchet_polyhedral-based_2013,grauer-gray_auto-tuning_2012}, on both backends, as shown in Table~\ref{tab:gpu-workloads}.
For all experiments, we selected input parameters (e.g., batch size, input image size, token length) to match common ML inference scenarios and ensure tractable simulation times.

The following sections present detailed experimental results:
Section~\ref{sec:gpu-experiments} examines GPU cache hierarchies, while Section~\ref{experiment-scalesim} explores dataflow-driven memory optimization in systolic arrays.
Together, these studies demonstrate GainSight's ability to inform lifetime-aware memory design across diverse accelerator platforms.

\subsection{Profiling GPU Workloads}\label{sec:gpu-experiments}

This case study demonstrates GainSight's lifetime-aware design methodology by evaluating GPU memory hierarchies using our Accel-Sim backend.
We analyze data lifetime patterns across L1 and L2 caches for representative AI workloads to evaluate the suitability of different memory devices for these caches and to quantify the benefits of heterogeneous memory compositions.

\subsubsection{Experimental Setup}

The workloads evaluated in this study (Table~\ref{tab:gpu-workloads}) span transformer-based language models from MLPerf Inference and computer vision benchmarks from PolyBench~\cite{pouchet_polyhedral-based_2013,grauer-gray_auto-tuning_2012}.
They represent the memory-intensive AI inference patterns identified in Section~\ref{sec:intro} that exhibit data transience characteristics, motivating heterogeneous memory architectures.

Text-based transformer workloads were configured with 7-word input prompts and 20-token generation sequences, while vision workloads processed $224 \times 224$ pixel images with batch size 1.
These configurations maintain statistical representativeness of real-world memory access patterns while ensuring tractable simulation runtimes through our kernel sampling methodology.
Full data and visualizations are available at \url{https://gainsight.stanford.edu/visualization/index.html} as part of this paper's artifacts.

\subsubsection{Data Lifetime Characterization}

Figure~\ref{fig:tableau} demonstrates the results of GainSight's data lifetime analysis and per-device performance evaluation for each workload running on the GPU backend.
The scatter plots project area and energy requirements for L1/L2 caches if these caches were implemented with each individual memory device. 
The histograms display data lifetime distributions for direct correlation with device retention specifications.
Vertical reference lines indicate Si-GCRAM (\SI{1}{\micro\second}) and Hybrid-GCRAM (\SI{10}{\micro\second}) retention times~\cite{liu_design_2024}.
The \si{\milli\second}-scale retention of eDRAM is not shown for conciseness.

Lifetime analysis reveals substantial fractions of accesses compatible with refresh-free operation on StRAM devices. Specifically, 64.3\% of L1 and 18.4\% of L2 accesses fall within Si-GCRAM's \SI{1}{\micro\second} retention, while 97.9\% of L1 and 52.0\% of L2 accesses fit Hybrid-GCRAM's \SI{10}{\micro\second} capability~\cite{liu_design_2024}.
Conversely, significant portions exhibit longer lifetimes (up to \SI{70}{\micro\second}), requiring heterogeneous architectures with longer-retention devices such as eDRAM or SRAM.

Kernel-level analysis reveals that short-lived data predominantly originate from linear algebra operations like GEMM, while longer data lifetimes emerge during memory transfers and nonlinear computations, aligning with Section~\ref{sec:intro}'s architectural motivations.

\begin{takeaway}
  Data transience is pervasive across the GPU cache hierarchy, with up to 97.9\% of L1 accesses and 52.0\% of L2 accesses fitting within StRAM retention times, enabling refresh-free operations for non-SRAM, heterogeneous memory architectures.
\end{takeaway}

\subsubsection{Optimal Heterogeneous Memory Configurations}

\begin{figure}[tb]
  \centering
  \includesvg[width=\linewidth]{imgs/gpu_hetero.svg}
  \caption{Optimal heterogeneous memory configurations showing capacity proportions of Si-GCRAM, Hybrid-GCRAM, and eDRAM, provisioned based on lifetime profiles for the GPU backend across various workloads.}
  \label{fig:composition}
\end{figure}

Leveraging the observed lifetime distributions, GainSight's frontend generates optimal memory compositions that minimize active energy while ensuring refresh-free operation across Si-GCRAM, Hybrid-GCRAM, and eDRAM technologies.
The optimization strategy systematically assigns data to devices with minimal necessary retention times, as detailed in Section~\ref{sec:frontend}, prioritizing energy-efficient short-retention memories while guaranteeing data integrity.

Figure~\ref{fig:composition} illustrates recommended heterogeneous compositions across all workloads in Table~\ref{tab:gpu-workloads} and cache levels.
Most AI/ML workloads benefit from Si-GCRAM-dominated L1 caches (>90\% of total capacity) with minimal longer-retention allocations. 
However, transformer models like Llama variants~\cite{grattafiori_llama_2024} require substantial longer-retention capacity for KV cache storage.
L2 caches exhibit more varied lifetime patterns necessitating balanced multi-technology allocations.

\begin{figure}[t]
  \centering
  \includegraphics[width=\linewidth]{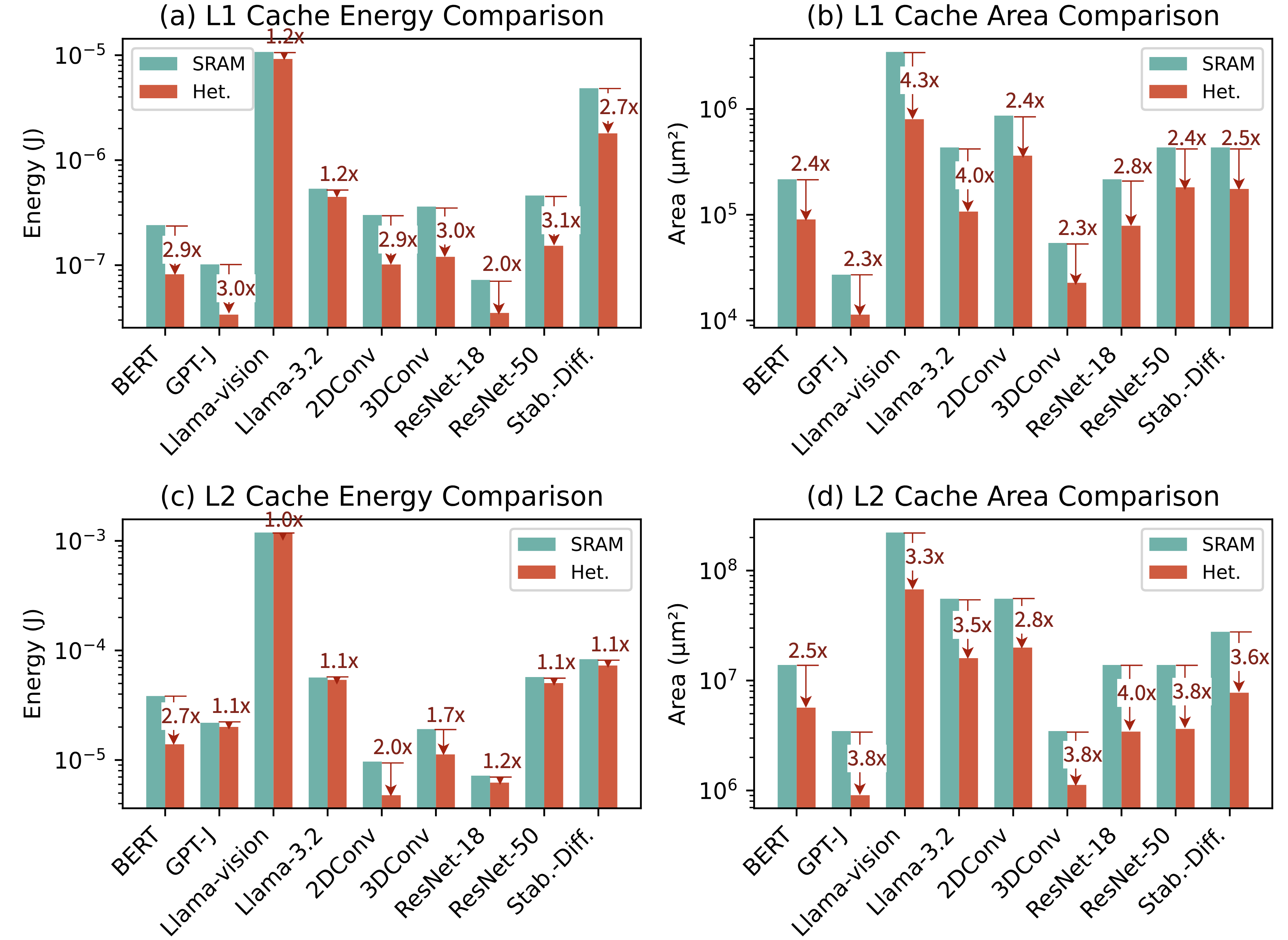}
  \caption{Comparison of area and energy savings of heterogeneous L1/L2 caches against monolithic SRAM designs across GPU workloads.}
  \label{fig:heterogeneous}
\end{figure}

Lifetime-optimized configurations achieve substantial improvements over monolithic SRAM designs (Figure~\ref{fig:heterogeneous}).
Si-GCRAM-dominant L1 caches realize an average 1.89\texttimes{} active energy savings, while Hybrid-GCRAM-integrated L2 caches achieve an average of 1.3\texttimes{} energy reductions.
A consistent memory array area improvement of 3\texttimes{} is observed across all workloads compared to baseline SRAM, demonstrating the superior density that a heterogeneous mix of non-SRAM StRAM devices can provide.

These results validate GainSight's core methodology that systematic lifetime analysis enables substantial improvements in memory system efficiency by exploiting the mismatch between data transience and SRAM overprovisioning identified in Section~\ref{sec:intro}.

\begin{takeaway}
  GainSight's lifetime-aware optimization enables heterogeneous memory architectures that achieve meaningful active energy reductions (up to 3.1\texttimes{}) and area savings (up to 4.3\texttimes{}) by systematically matching workload data lifetimes with optimal memory device retention.
\end{takeaway}

\subsection{Profiling Workloads on Systolic Arrays}\label{experiment-scalesim}

This case study demonstrates GainSight's versatility across architectures through systematic evaluation of systolic array memory hierarchies using our SCALE-Sim backend.
Systolic arrays represent a different paradigm from GPUs, with dedicated scratchpads and deterministic dataflows that facilitate controlled lifetime-retention analysis~\cite{tpu_paper_jouppi}.

\subsubsection{Experimental Configuration}

We evaluated three representative AI workloads, ResNet-50~\cite{he_deep_2015}, BERT~\cite{devlin_bert_2019}, and Llama-3-1B~\cite{touvron_llama_2023}, across PE array configurations ranging from $32 \times 32$ to $256 \times 256$ elements.
Each workload's GEMM operations were analyzed under three canonical dataflows: input stationary (\textit{is}), weight stationary (\textit{ws}), and output stationary (\textit{os}).
The accelerator configuration included \SI{4}{\kilo\byte} input and weight buffers with an \SI{8}{\kilo\byte} output buffer, representing typical edge-oriented systolic array designs.

BERT and Llama used 256-token sequences; ResNet-50 processed $224 \times 224$ images with single batch execution.
These configurations focus on GEMM operations dominating AI inference, while nonlinear operations represent future extensions.
For simplicity, we focus on ResNet-50 results in this section, with full experimental datasets available as part of this paper's artifacts.

\subsubsection{Dataflow-Dependent Lifetime Patterns}
\label{experiment-scalesim-results}

\begin{figure}[t]
  \centering
  \includegraphics[width=\linewidth]{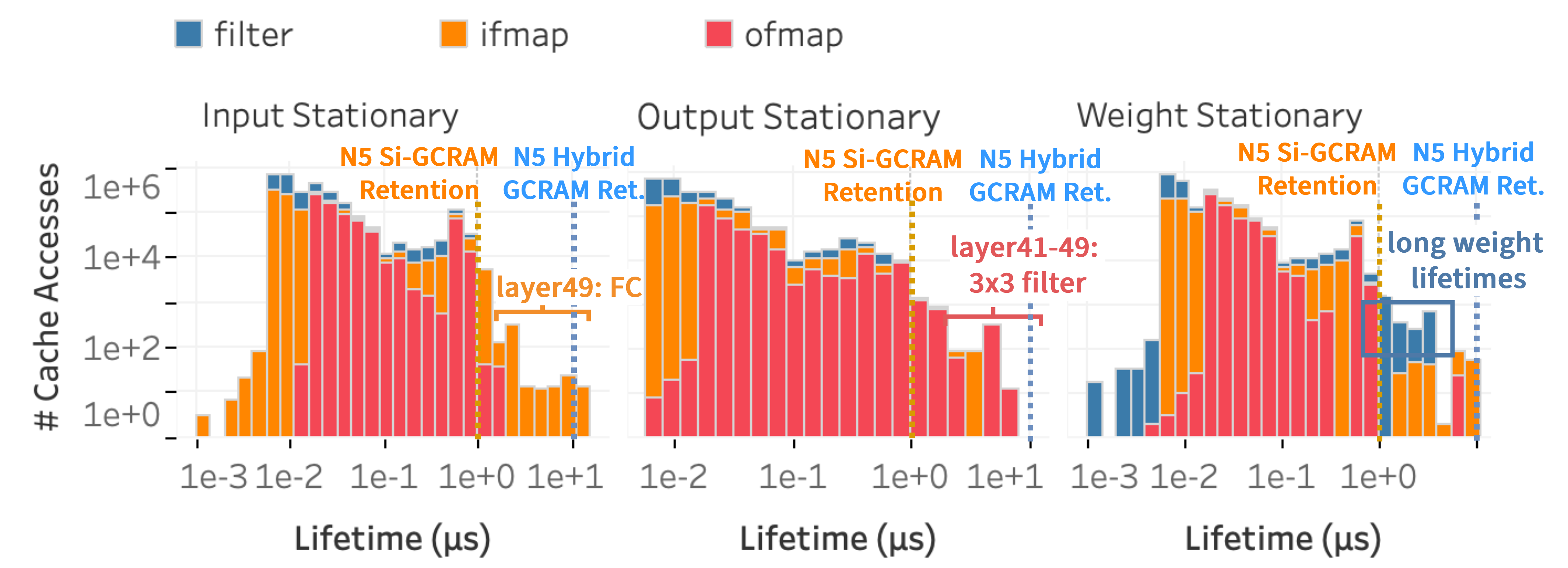}
  \caption{Data lifetime distributions for ResNet-50 on a $256 \times 256$ systolic array across three dataflows.}
  \label{fig:dataflow}
\end{figure}

Figure~\ref{fig:dataflow} presents lifetime distributions for ResNet-50 on a $256 \times 256$ array across three dataflows, demonstrating how mapping decisions influence memory patterns and data lifetimes.
Vertical reference lines indicate Si-GCRAM and Hybrid-GCRAM retention times.
It can be observed that at least 79.01\% of data accesses across input, output, and weight buffers exhibit lifetimes below \SI{1}{\micro\second}, making them suitable for refresh-free storage in 5nm Si-GCRAM.

The histograms show the heterogeneous nature of data lifetimes across buffers and dataflows.
Short-lived data emerges from rapid streaming between scratchpads and PE arrays or memory, while long-lived data arises during inactivity periods following PE array saturation in the buffers for the stationary component of the dataflow.
This heterogeneity aligns with GPU observations from Section~\ref{sec:gpu-experiments}, suggesting common patterns across accelerator architectures.

Analysis across dataflows reveals two key insights.
First, output feature map (\textit{ofmap}) data consistently exhibits relatively short lifetimes across dataflows, with nearly all data compatible with Hybrid-GCRAM refresh-free operation.
This behavior stems from the tight coupling between PE array timing and output feature map generation.
Second, dataflow selection primarily affects the maximum lifetime of stationary data structures, as algorithmic stationarity directly correlates with buffer data lifetimes.

\begin{takeaway}
Independent of dataflow, \textit{ofmap} data demonstrates consistently short lifetimes suitable for StRAM integration, while \textit{ifmap} and weight data exhibit broader distributions with dataflow-dependent upper bounds.
\end{takeaway}

\subsubsection{Heterogeneous Scratchpad Optimization}

Building upon lifetime characterizations, GainSight generates optimal heterogeneous scratchpad compositions for both stationary and non-stationary data types without refresh overheads (Figure~\ref{fig:scalesim_hetero_comp}).
The vast majority of systolic array scratchpad data across dataflows exhibits short-enough lifetimes compatible with Si-GCRAM, requiring minimal higher-retention Hybrid-GCRAM or eDRAM capacity.

\begin{figure}[t]
  \centering
  \includesvg[width=\linewidth]{imgs/scalesim_hetero_comp.svg}
  \caption{Optimal heterogeneous memory configurations showing capacity proportions of Si-GCRAM, Hybrid-GCRAM, and eDRAM, provisioned based on lifetime profiles for the systolic array backend on ResNet-50.}
  \label{fig:scalesim_hetero_comp}
\end{figure}

\begin{figure}[t]
  \centering
  \includegraphics[width=\linewidth]{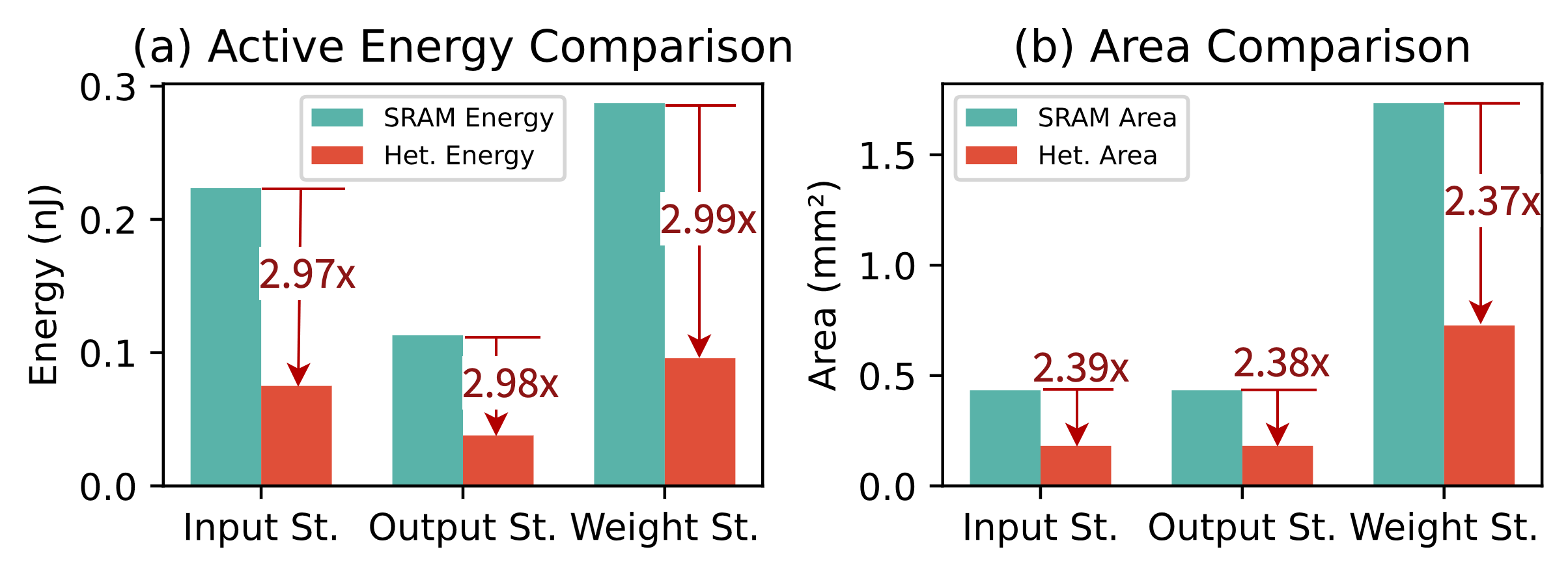}
  \caption{Comparison of area and energy savings of a heterogenous scratchpad memory against monolithic SRAM designs on ResNet-50 across three dataflows: Input St(ationary), Output St(ationary), and Weight St(ationary).}
  \label{fig:scalesim_hetero}
\end{figure}

Lifetime-optimized heterogeneous on-chip memory designs achieve substantial improvements over uniform SRAM scratchpads.
Figure~\ref{fig:scalesim_hetero} quantifies 3\texttimes{} active energy savings and 2.4\texttimes{} area savings across all dataflows.

These consistent gains regardless of dataflow reflect the deterministic and predictable nature of systolic array buffer access patterns, where data movement follows well-defined dataflow protocols.
These consistent benefits contrast with the more variable improvements observed in GPU experiments, highlighting how architectural determinism in systolic arrays enables more predictable lifetime-driven optimization outcomes compared to the dynamic control flow complexity in SIMD GPU kernels.

\subsubsection{PE Array Scaling Analysis}

An important design consideration in systolic arrays is the size of the processing element (PE) array, which influences data lifetime metrics and scratchpad interactions beyond just area and energy implications.

Figure~\ref{fig:lifetimes_vs_array} demonstrates systematic relationships between PE array size and lifetime characteristics.

\begin{figure}[t]
  \centering
  \includesvg[width=\linewidth]{imgs/array_size.svg}
  \caption{Average and maximum data lifetimes across different PE array sizes for ResNet-50 on the SCALE-Sim backend.}
  \label{fig:lifetimes_vs_array}
\end{figure}

Larger PE arrays not only incrementally reduce average data lifetimes due to faster data consumption but also narrow the variance of lifetime distributions, where the maximum lifetime of stationary data types decreases significantly.

This systematic reduction in lifetime variance creates opportunities for applying StRAM devices that more aggressively trade off retention time for other benefits such as energy efficiency. 
This becomes especially beneficial when outlier data lifetimes would otherwise significantly contribute to overall refresh overhead.
This observation demonstrates that GainSight's utility extends beyond informing on-chip memory provisioning to broader architectural design decisions.

\begin{takeaway}
  Larger systolic arrays narrow the variance of data lifetimes and eliminate long-lived outliers, creating further StRAM integration opportunities.
\end{takeaway}

\section{Discussion}\label{sec:conclusion}

We have presented GainSight, the first profiling and design exploration framework that correlates dynamic workload lifetime statistics with memory device characteristics to enable optimal heterogeneous memory compositions in AI accelerators.
This section discusses GainSight's broader applicability beyond AI accelerators, examines the future scalability of StRAM technologies, and outlines directions for extending the framework to encompass long-term memory devices.

\subsection{Data Lifetime Heterogeneity Beyond AI Workloads}

GainSight's lifetime-aware design methodology extends naturally beyond AI accelerators to other domains exhibiting heterogeneous data lifetimes and pervasive short-livedness.

Server workloads exemplify this potential.
HTTP buffers and session state demonstrate \si{\micro\second}-scale lifetimes ideal for the short-term RAM (StRAM) devices evaluated by GainSight, while cached content suits long-term RAM (LtRAM) technologies in Table~\ref{tab:memory_types}.
Database systems benefit similarly, with transaction logs and query results exhibiting transient patterns favoring StRAM.

Scientific and high-performance computing demonstrates comparable patterns.
Simulations generate massive intermediate arrays discarded after convergence -- classic StRAM applications -- while mesh topology remains static throughout runs.
Multimedia pipelines exhibit similar characteristics with brief frame buffers during compression.

These examples underscore the universality of lifetime-aware design principles that GainSight helps establish.

\subsection{Future of GCRAM Scaling}

Given our focus on GCRAM and other StRAM devices, it is important to consider the future of these technologies to ensure they do not fall into the same scaling trap as SRAM.
For GCRAM devices, prior work~\cite{liu_design_2024} highlights scaling challenges.
Decreasing process nodes increase leakage power and reduce retention time in both Si-GCRAM and Hybrid-GCRAM, even as technologies such as Gate-All-Around (GAA) FETs~\cite{yeap_2nm_2024} can reduce the off-current of transistors due to enhanced gate control compared to FinFETs.
However, we believe that GCRAM will continue to be a viable option for short-term on-chip memory arrays.
Notably, the expected improvement in operating frequency with smaller process nodes mitigates retention requirements, as increasing frequencies allow for more instructions to be completed within a given time frame, reducing real-world lifetimes.
Additionally, various circuit-level techniques are available to further enhance retention in advanced technology nodes~\cite{yigit_128-kbit_2023,harel_16-kb_2024}.

\subsection{Evaluating Long-Term RAM Devices}

GainSight currently focuses on on-chip StRAM technologies, specifically GCRAM variants and embedded DRAM.
This reflects the significant opportunities for optimizing short-lived data patterns in AI workloads and other data-intensive applications, as demonstrated by our experimental results.
While LtRAM devices like RRAM and MRAM offer complementary potential~\cite{rram-dnn, li_combating_2012}, they are excluded because the data transience and high write intensities in this study's target workloads misalign with LtRAM characteristics.
Moreover, effective LtRAM evaluation requires comprehensive endurance modeling that accounts for write frequency, spatial patterns, and temporal clustering over extended operational periods.

Nonetheless, GainSight's architecture- and device-agnostic infrastructure provides a robust foundation for future extensions to include LtRAM technologies.
Subsequent iterations can incorporate LtRAM device models and analytical layers for endurance-aware evaluations.
This provides a promising direction for extending heterogeneous memory design across the full spectrum of StRAM and LtRAM technologies.

\section{Conclusion}

As SRAM scaling plateaus while AI workloads demand exponentially growing memory capacity, GainSight establishes the first systematic methodology for exploiting the pervasive data transience in these workloads through lifetime-aware heterogeneous on-chip memory design.
Our framework bridges the gap between dynamic workload characteristics and emerging short-term memory technologies, transforming accelerator memory design from intuition-driven to profile-guided.

GainSight's three synergistic contributions -- heterogeneous memory composition, unified profiling and evaluation framework, and case studies -- provide concrete foundations for the next generation of AI accelerator memory systems.
By establishing data lifetime as a first-class design constraint and delivering practical tooling for heterogeneous on-chip memory composition, GainSight enables the architecture community to reimagine on-chip memory as a differentiated resource optimized for data transience.

\begin{acks}
  This project is funded by the Microelectronics Commons Program, a DoD initiative.
\end{acks}
\bibliographystyle{ACM-Reference-Format}

\bibliography{main}

\end{document}